\documentclass[aps,preprint]{revtex4}%
\usepackage{amsfonts}
\usepackage{amsmath}
\usepackage{amssymb}
\usepackage{graphicx}%
\begin{document}
\title[ ]{Classical Physics of Thermal Scalar Radiation in Two Spacetime Dimensions}
\author{Timothy H. Boyer}
\affiliation{Department of Physics, City College of the City University of New York, New
York, New York 10031}

\begin{abstract}
Thermal scalar radiation in two spacetime dimensions is treated within
relativistic classical physics. \ Part I involves an inertial frame where are
given the analogues both of Boltzmann's derivation of the \ Stefan-Boltzmann
law and also Wien's derivation of the displacement theorem using the
$\sigma_{ltU^{-1}}$-scaling of relativistic radiation theory. \ Next the
spectrum of classical scalar zero-point radiation in an inertial frame is
derived both from $\sigma_{ltU^{-1}}$-scale invariance and from Lorentz
invariance. \ Part II involves the behavior of thermal radiation in a
coordinate frame undergoing (relativistic) constant acceleration, a Rindler
frame. \ The radiation normal modes in a Rindler frame are obtained. \ The
classical zero-point radiation of inertial frames is transformed over to the
coordinates of a Rindler frame. \ Although for zero-point radiation the
two-field correlation function at different spatial points at a single time is
the same between inertial and Rindler frames, the correlation function at two
different times at a single Rindler spatial coordinate is different, and has a
natural extension to non-zero temperature. \ The thermal spectrum in the
Rindler frame is then transferred back to an inertial frame, giving the
familiar Planck spectrum. \ 

\end{abstract}
\maketitle

\section{Introduction}

Thermal radiation holds an unusual place in physics. \ Although thermal
radiation provides a simple system for the discussion of thermodynamics, the
determination of the spectrum of thermal radiation is often regarded as an
intractable problem within classical physics. \ Here we provide a discussion
of thermal radiation within classical physics which is different from the
historical treatments. \ Rather than attempting to determine the thermal
spectrum from use of nonrelativistic classical statistical mechanics or from
scattering by nonrelativistic mechanical systems, we use the scaling
symmetries of relativistic classical radiation both in inertial frames and in
(relativistic) accelerating coordinate frames. The Planck spectrum arises
naturally from ideas of classical radiation equilibrium. \ The analysis also
suggests that relativistic physics is important in problems of classical
radiation equilibrium.

The present article is at the interface between pedagogy and research. \ In
the first part, we review some of the ideas of thermal radiation in inertial
frames which appear in text books of thermodynamics\cite{Morse} and of modern
physics\cite{Eisberg} but in the simpler context of scalar radiation in two
spacetime dimensions. \ We give the analogues of Boltzmann's derivation of the
Stefan-Boltzmann law\cite{Morse78-79} and Wein's derivation of the
displacement theorem\cite{Lavenda67-70}. \ However, the analysis also
emphasizes aspects which are rarely treated in the text books, such as the
$\sigma_{ltU^{-1}}$-scale invariance of relativistic radiation theory, and
also the Lorentz invariance of the radiation spectrum which assigns to each
normal mode an energy equal to a constant times the frequency. \ In the second
part we introduce a Rindler coordinate frame corresponding to a system
accelerating relative to an inertial frame. \ Now a Rindler frame is a more
complicated system than an inertial frame and is usually introduced in
connection with general relativity.\cite{Schutz} \ However, a Rindler frame
actually involves no more than the ideas of special relativity.\cite{Rindler}
\ Since classical radiation transforms as a tensor, we expect that classical
zero-point radiation will reappear in any other coordinate frame, including a
Rindler frame. \ Scalar radiation has trivial tensor transformations and so is
easily transferred to the coordinates of the Rindler frame. \ The radiation
normal modes in the Rindler frame are introduced as a natural generalization
of the radiation modes in an inertial frame. \ In a Rindler frame, there is a
natural extension of the radiation correlation function from zero-point
radiation over to thermal radiation. \ Finally we show how to bring the
information about the spectrum of thermal radiation in a Rindler frame back to
an inertial frame.

The analysis presented here in the simple case of scalar radiation and two
spacetime dimensions involves only elementary functions\cite{modified} and
extensions of familiar ideas. \ Nevertheless, the conclusions are radically
different from what appears in current text books of modern physics. \ The
present analysis emphasizes the relativistic nature of the thermal radiation
of a relativistic field.

\section{Part I: Thermal Radiation in an Inertial Frame}

\subsection{Simplified Model Theory}

\subsubsection{Basic Assumptions}

Our discussion of radiation equilibrium begins with certain fundamental
assumptions based upon experimental observations. \ 1) We assume that we are
dealing with a relativistic radiation field so that the speed of light in
vacuum $c$ enters as a fundamental constant. \ 2) We assume that radiation
equilibrium exists in the form corresponding to Stefan's law for
electromagnetic radiation, $u_{T}=\sigma_{S}T^{4},$ where $u_{T}$ is the
thermal energy density per unit volume, $T$ is the absolute temperature, and
$\sigma_{S}$ (Stefan's constant) is a fundamental constant. \ 3) We assume
there is random radiation present at the absolute zero of temperature. \ 4) We
assume that thermal equilibrium exists in a uniformly accelerated coordinate
frame or (by the equivalence principle) in a gravitational field.

The assumptions we have made are taken from experimental observations.
\ Maxwell's equations for the electromagnetic field were developed in the
1860s based upon the electromagnetic observations of Ampere, Faraday, and
their contemporaries. \ These equations are relativistically invariant and
contain the speed of light in vacuum $c$. \ The development of thermodynamics
also occurred during the nineteenth century and Stefan's
observations\cite{Stefan} regarding thermal radiation were made in the 1870's.
\ The idea of zero-point radiation arose during the twentieth century and the
experimental observations of Spaarnay, Lamoreau and others\cite{Spaarnay}
confirm the calculations based upon zero-point radiation.\cite{CasimirF}
\ Finally, it is a natural extension of our thermodynamics observations to
assume that thermodynamics ideas hold in gravitational fields. \ Such ideas
were used in the 1870s by Boltzmann in his derivation of the Maxwell velocity
distribution for thermal particle velocities.\cite{BoltzMax}

\subsubsection{Scalar Radiation in Two Spacetime Dimensions}

In this article, we wish to simplify the mathematics as much as possible while
exploring the implications of thermal radiation. \ Therefore we will not treat
the electromagnetic field in four spacetime dimensions (three space and one
time dimension), but rather will discuss a model calculation involving scalar
radiation in two spacetime dimensions (one space and one time
dimension).\cite{4dim} \ Instead of the six fields $E_{x},E_{y},E_{z,}%
B_{x},B_{y},B_{z}$ of electromagnetic theory satisfying Maxwell's equations
in\thinspace\ ($ct,x,y,z)$, we will consider only one relativistic scalar
field $\phi$ which is a function of ($ct,x)$ in an inertial frame with
spacetime metric $ds^{2}=g_{\mu\nu}dx^{\mu}dx^{\nu}$ where the indices $\mu$
and $\nu$ run over 0 and 1, $x^{0}=ct,$ $x^{1}=x,$
\begin{equation}
ds^{2}=c^{2}dt^{2}-dx^{2}%
\end{equation}
\ The behavior of the field $\phi$ follows from the Lagrangian density
$\mathcal{L}=(1/8\pi)\partial^{\mu}\phi\partial_{\mu}\phi$ corresponding
to\cite{Goldstein}%
\begin{equation}
\mathcal{L}=\frac{1}{8\pi}\left[  \frac{1}{c^{2}}\left(  \frac{\partial\phi
}{\partial t}\right)  ^{2}-\left(  \frac{\partial\phi}{\partial x}\right)
^{2}\right]
\end{equation}
and the wave equation $\partial_{\mu}[\partial\mathcal{L}/\partial
(\partial_{\mu}\phi)]=0$ for the field is
\begin{equation}
\frac{1}{c^{2}}\left(  \frac{\partial^{2}\phi}{\partial t^{2}}\right)
-\left(  \frac{\partial^{2}\phi}{\partial x^{2}}\right)  =0
\end{equation}
The associated stress-energy-momentum tensor density is given by
$\mathcal{T}^{\mu\nu}=[\partial\mathcal{L}/\partial(\partial_{\mu}%
\phi)]\partial^{\nu}\phi-g^{\mu\nu}\mathcal{L}$ so that the energy density
$u=\mathcal{T}$ $^{00}$ is
\begin{equation}
\mathcal{T}^{00}=-\mathcal{T}^{11}=\frac{1}{8\pi}\left[  \frac{1}{c^{2}%
}\left(  \frac{\partial\phi}{\partial t}\right)  ^{2}+\left(  \frac
{\partial\phi}{\partial x}\right)  ^{2}\right]
\end{equation}
and the momentum density is
\begin{equation}
\mathcal{T}^{01}=\mathcal{T}^{10}=\frac{1}{4\pi c}\frac{\partial\phi}{\partial
t}\frac{\partial\phi}{\partial x}%
\end{equation}
The energy $U$ in the field in a one-dimensional box extending from $x=a$ to
$x=b$ follows as%
\begin{equation}
U=\int_{a}^{b}dx\frac{1}{8\pi}\left[  \frac{1}{c^{2}}\left(  \frac
{\partial\phi}{\partial t}\right)  ^{2}+\left(  \frac{\partial\phi}{\partial
x}\right)  ^{2}\right]
\end{equation}

\subsection{Thermodynamics of Radiation in a One-Dimensional Box}

The thermodynamics of radiation in a box (in this case a box with one spatial
dimension) can be found in a traditional fashion from
\begin{equation}
TdS=dU_{T}+p_{T}dV
\end{equation}
In the present case of one spatial dimension, the length $L$ will replace the
volume $V$. \ Also, for a large box, we can neglect the (Casimir)
effects\cite{CasimirF} due to the discrete nature of the normal modes so that
the energy density in an inertial frame is uniform throughout the box and the
energy density is a function of temperature alone so that the thermodynamic
equation (7) becomes
\begin{equation}
TdS=d(u_{T}L)+p_{T}dL=Ldu_{T}+(u_{T}+p_{T})dL
\end{equation}
From the tensor equation (4) above, we see that the energy density and stress
tensor density are equal in magnitude $T^{00}=-T^{11}$ so that the pressure
$p_{T}=u_{T},$ and Eq. (8) becomes%
\begin{equation}
TdS=T\left(  \frac{\partial S}{\partial T}\right)  _{L}dT+T\left(
\frac{\partial S}{\partial L}\right)  _{T}dL=L\frac{du_{T}}{dT}dT+2u_{T}dL
\end{equation}
This gives us%
\begin{equation}
\left(  \frac{\partial S}{\partial T}\right)  _{L}=\frac{L}{T}\frac{du_{T}%
}{dT}%
\end{equation}
and%
\begin{equation}
\left(  \frac{\partial S}{\partial L}\right)  _{T}=\frac{2u_{T}}{T}%
\end{equation}
The equality of the mixed partials $\partial^{2}S/(\partial T\partial
L)=\partial^{2}S/(\partial L\partial T)$ then implies%
\begin{equation}
\frac{\partial}{\partial L}\left(  \frac{L}{T}\frac{du_{T}}{dT}\right)
=\frac{\partial}{\partial T}\left(  \frac{2u_{T}}{T}\right)
\end{equation}
or%
\begin{equation}
\frac{1}{T}\frac{du_{T}}{dT}=\frac{2}{T}\frac{du_{T}}{dT}-\frac{2u_{T}}{T^{2}}%
\end{equation}
with solution%
\begin{equation}
u_{T}=\sigma_{2D}T^{2}\text{ \ \ \ \ }U_{T}=\sigma_{2D}T^{2}L
\end{equation}
The entropy $S$ follows from Eq. (11) as%
\begin{equation}
S=2\sigma_{2D}TL
\end{equation}
Thus we have found for a one-dimensional box the thermodynamic expressions
exactly analogous to Boltzmann's familiar expressions for a three-dimensional
box. \ The constant $\sigma_{2D}$ is analogous to Stefan's constant
$\sigma_{S}$ for electromagnetic radiation in three-dimensional space.

\subsection{Scaling and Fundamental Constants}

\subsubsection{$\sigma_{ltU^{-1}}$-Scaling for Any Radiation Theory Allowing
Thermal Equilibrium}

Classical mechanics has no fundamental constants and accordingly allows
separate scalings in length, time, and energy. \ Thus, for example, in
principle, any classical mechanical system can be made twice as large, can
perform at three times the speed, and involve four times the energy.
\ However, a classical theory which contains fundamental constants has its
scaling constrained by these fundamental constants. \ This situation is
familiar in connection with the fundamental constant $c$ which enters
relativistic systems$.$ \ The fundamental constant $c$ involves length $l$
divided by time $t$ and so couples together the scalings of length and time so
as to preserve the value of the fundamental constant $c$. \ In the paragraph
above, we have seen that if thermal radiation equilibrium exists, then there
must be a fundamental constant corresponding to Stefan's constant $\sigma
_{S}\,,\ $which in our two-dimensional case is $\sigma_{2D}.$ \ The
fundamental constant $\sigma_{S}$ (or in our case $\sigma_{2D})$ couples
together energy and length appearing in the thermal energy density $u_{T}$ and
the thermal energy $k_{B}T$. \ Thus for any relativistic classical radiation
theory which allows thermal equilibrium, the fundamental constants allow only
those scalings which couple together length, time, and energy: length scales
as $l\rightarrow l^{\prime}=\sigma l,$ time scales as $t\rightarrow t^{\prime
}=\sigma t,$ and energy scales as $U\rightarrow U^{\prime}=U/\sigma,$ for
$\sigma>0.$ \ Thus we say that the theory satisfies a $\sigma_{ltU^{-1}}%
$-scale invariance.\cite{Scale} \ Under such a scaling, the fundamental
constants $c$ and $\sigma_{2D}$ are carried into themselves, the wave equation
(3) is carried into itself, and solutions of the differential equation are
carried into solutions.

\subsubsection{$\sigma_{ltU^{-1}}$-Scaling and Adiabatic Change in an Inertial
Frame}

For a relativistic classical radiation theory which contains no fundamental
lengths, a $\sigma_{ltU^{-1}}$-scale change is equivalent to an adiabatic
change. \ Thus, for example, if the standards of measurement for length, time,
and energy undergo a $\sigma_{ltU^{-1}}$-scale change, then a thermodynamic
system will be interpreted as having a new temperature $T/\sigma$, a new box
length $\sigma L$, and new energy $U/\sigma$. \ However, the entropy $S$ of
the system in Eq. (15) is invariant under such a scale change. \ These changes
are exactly the same as though one had carried out an adiabatic compression or
expansion of the radiation in the box. \ Under an adiabatic change, there is
no change in the entropy of the system. \ Accordingly, equations (7) and (8)
become
\begin{equation}
0=dU_{T}+p_{T}dL=dU_{T}+\frac{U_{T}}{L}dL
\end{equation}
or
\begin{equation}
U_{T}=\frac{const}{L}%
\end{equation}
during an adiabatic change. \ But this equation (17) is invariant under a
$\sigma_{ltU^{-1}}$-scale change since $U_{T}\rightarrow U_{T}/\sigma$ while
$L\rightarrow\sigma L.$ \ Indeed the thermodynamics equations (14) and (15)
transform under adiabatic change just as though the standards of measurement
for length, time, and energy had undergone a $\sigma_{ltU^{-1}}$-scale change. \ 

Although the theoretical calculation of Stefan's law\cite{Morse78-79}
(comparable to our calculation here) was carried through by Boltzmann in the
1880s and is thoroughly familiar to many students, the remarks about
$\sigma_{ltU^{-1}}$-scaling are likely to be unfamiliar. \ However, they will
be useful in our later work. \ 

\subsection{Radiation Spectrum in a Box}

\subsubsection{Radiation Normal Modes}

The radiation in a box can be described by a complete set of either standing
waves or running waves with appropriate wave vectors$\mathbf{.}$ $\ $In the
present case, we will choose standing wave solutions which vanish at the walls
$x=a$ and $x=b$ of the box (Dirichlet boundary conditions) so that a
normalized normal mode can be written as
\begin{equation}
\phi_{n}(ct,x)=f_{n}\left(  \frac{2}{b-a}\right)  ^{1/2}\sin\left[  \frac
{n\pi}{b-a}(x-a)\right]  \cos\left[  \frac{n\pi}{b-a}ct+\theta_{n}\right]
\end{equation}
\ Then the radiation field in the box can be written as a sum over all the
normal modes
\begin{equation}
\phi(ct,x)=\sum_{n=1}^{\infty}f_{n}\left(  \frac{2}{b-a}\right)  ^{1/2}%
\sin\left[  \frac{n\pi}{b-a}(x-a)\right]  \cos\left[  \frac{n\pi}%
{b-a}ct+\theta_{n}\right]
\end{equation}
where $\theta_{n}$ is an appropriate phase. \ From Eq. (5), we find that each
mode $\phi_{n}(ct,x)$ has the time-average spatial energy density%
\begin{align}
u_{n}(x)  &  =\left\langle \frac{1}{8\pi}\left[  \frac{1}{c^{2}}\left(
\frac{\partial\phi_{n}}{\partial t}\right)  ^{2}+\left(  \frac{\partial
\phi_{n}}{\partial x}\right)  ^{2}\right]  \right\rangle _{time}\nonumber\\
&  =\frac{1}{8\pi}\left(  \frac{n\pi}{b-a}\right)  ^{2}f_{n}^{2}\frac{2}%
{b-a}\{\sin^{2}\left[  \frac{n\pi}{b-a}(x-a)\right]  \left\langle \sin
^{2}\left[  \frac{n\pi}{b-a}ct+\theta_{n}\right]  \right\rangle _{time}%
\nonumber\\
&  +\cos^{2}\left[  \frac{n\pi}{b-a}(x-a)\right]  \left\langle \cos^{2}\left[
\frac{n\pi}{b-a}ct+\theta_{n}\right]  \right\rangle _{time}\}\nonumber\\
&  =\frac{1}{8\pi}\left(  \frac{n\pi}{b-a}\right)  ^{2}\frac{f_{n}^{2}}{b-a}%
\end{align}
which is uniform in space. The total mode energy $U_{n}$ found by integrating
over the length is given by%
\begin{equation}
U_{n}=\frac{1}{8\pi}\left(  \frac{n\pi}{b-a}\right)  ^{2}f_{n}^{2}%
\end{equation}

\subsubsection{Two-Point Correlation Function for Random Radiation}

Coherent radiation involves fixed phase relations $\theta_{n}-\theta
_{n^{\prime}}$ between the various modes $\phi_{n}$ and $\phi_{n^{\prime}}$
which are used to decompose a radiation pattern. \ Random radiation, such as
is involved in thermal radiation, involves the opposite situation. \ Random
radiation can be written in the form of Eq. (19) where the phases $\theta_{n}$
are randomly distributed on the interval $[0,2\pi)$ and are independently
distributed for each $n\mathbf{.}$ \ It is convenient to characterize random
radiation by taking the two-point correlation function of the fields
$\left\langle \phi(ct,x)\phi(ct^{\prime},x^{\prime})\right\rangle $ obtained
by averaging over the random phases as%
\begin{equation}
\left\langle \cos\theta_{n}\cos\theta_{n^{\prime}}\right\rangle =\left\langle
\sin\theta_{n}\sin\theta_{n^{\prime}}\right\rangle =(1/2)\delta_{n,n^{\prime}}%
\end{equation}%
\begin{equation}
\left\langle \cos\theta_{n}\sin\theta_{n^{\prime}}\right\rangle =0
\end{equation}
\ The two-point correlation function for a general distribution of random
classical scalar waves is found by averaging over the random phases
$\theta_{n}$\cite{corr}%
\begin{align}
\left\langle \phi(ct,x)\phi(ct^{\prime},x^{\prime})\right\rangle  &
=\text{AV}\sum_{n=1}^{\infty}f_{n}\left(  \frac{2}{b-a}\right)  ^{1/2}%
\sin\left[  \frac{n\pi}{b-a}(x-a)\right]  \cos\left[  \frac{n\pi}%
{b-a}ct+\theta_{n}\right] \nonumber\\
&  \times\sum_{n^{\prime}=1}^{\infty}f_{n^{\prime}}\left(  \frac{2}%
{b-a}\right)  ^{1/2}\sin\left[  \frac{n^{\prime}\pi}{b-a}(x^{\prime
}-a)\right]  \cos\left[  \frac{n^{\prime}\pi}{b-a}ct^{\prime}+\theta
_{n^{\prime}}\right]  \text{AV}\nonumber\\
&  =\sum_{n=1}^{\infty}\frac{f_{n}^{2}}{b-a}\sin\left[  \frac{n\pi}%
{b-a}(x-a)\right]  \sin\left[  \frac{n\pi}{b-a}(x^{\prime}-a)\right]
\cos\left[  \frac{n\pi}{b-a}c(t-t^{\prime})\right]
\end{align}

Since we are not interested in the discrete mode structure of the box but
rather in the large-box limit, it is convenient to use the identity $2\sin
A\sin B=\cos(A-B)-\cos(A+B)$ to rewrite the correlation function in Eq. (24)
as%
\begin{align}
\left\langle \phi(ct,x)\phi(ct^{\prime},x^{\prime})\right\rangle  &  =\frac
{1}{2}\sum_{n=1}^{\infty}\frac{f_{n}^{2}}{b-a}\cos\left[  \frac{n\pi}%
{b-a}(x-x^{\prime})\right]  \cos\left[  \frac{n\pi}{b-a}c(t-t^{\prime})\right]
\nonumber\\
&  -\frac{1}{2}\sum_{n=1}^{\infty}\frac{f_{n}^{2}}{b-a}\cos\left[  \frac{n\pi
}{b-a}(x+x^{\prime}-2a)\right]  \cos\left[  \frac{n\pi}{b-a}c(t-t^{\prime
})\right]
\end{align}
We are interested only in points $x$ and $x^{\prime}$ which are far from the
edges of a large box, and so we will drop the term involving the large
separation $(x+x^{\prime}-2a)$ where the cosine function is oscillating very
rapidly. \ Also, for $b-a$ large (corresponding to a large box), we will
replace the sum by an integral $\sum\rightarrow\int dn,$ and will write
$k=n\pi/(b-a)$ so that%
\begin{align}
\left\langle \phi(ct,x)\phi(ct^{\prime},x^{\prime})\right\rangle  &  =\frac
{1}{2}\int_{0}^{\infty}dn\frac{f_{n}^{2}}{b-a}\cos\left[  \frac{n\pi}%
{b-a}(x-x^{\prime})\right]  \cos\left[  \frac{n\pi}{b-a}c(t-t^{\prime})\right]
\nonumber\\
&  =\frac{1}{2\pi}\int_{0}^{\infty}dkf^{2}(k)\cos\left[  k(x-x^{\prime
})\right]  \cos\left[  kc(t-t^{\prime})\right] \nonumber\\
&  =\frac{1}{4\pi}\int_{-\infty}^{\infty}dkf^{2}(|k|)\cos[k(x-x^{\prime
})-|k|c(t-t^{\prime})]
\end{align}
where in the last line we have used the identity $2\cos A\cos B=\cos
(A+B)+\cos(A-B)$ and have incorporated the two cosine terms by adding the
integral over negative values of $k$. \ The connection between the spectral
function $f^{2}(|k|)$ and the energy $U(k)$ of a normal mode of wave number
$k$ follows from Eq. (21) as
\begin{equation}
U(k)=\frac{1}{8\pi}k^{2}f^{2}(|k|)
\end{equation}

\subsection{Classical Zero-Point Radiation}

\subsubsection{Zero-Point Radiation in Classical Physics}

Experimental measurements\cite{Spaarnay} have confirmed that Casimir forces
can be described by random classical radiation\cite{CasimirF} at temperature
$T=0.$ \ Thus any theory of classical thermal radiation must also included
classical zero-point radiation as the limit when the temperature goes to
absolute zero. \ We have seen above that, for non-zero temperature, a
$\sigma_{ltU^{-1}}$-scale transformation carries a situation at temperature
$T$ into a situation at temperature $T/\sigma.$ \ At the absolute zero of
temperature, both $T$ and $T/\sigma$ are the same so that there can be no
change under rescaling; accordingly we expect $\sigma_{ltU^{-1}}$-scale
invariance for the random zero-point radiation. \ Thus the vacuum should not
have any finite-valued parameter associated with energy. \ This requirement of
$\sigma_{ltU^{-1}}$-scale invariance allows us to determine the spectrum of
classical zero-point radiation.

\subsubsection{Requirement of $\sigma_{ltU^{-1}}$-Scale Invariance of
Zero-Point Radiation in an Inertial Frame}

From the energy density equation (4) in two spacetime dimensions, we require
that ($\partial_{x}\phi)^{2}$ have the dimensions of energy divided by length.
\ Therefore $\phi^{2}$ has the dimensions of energy times length and so is
$\sigma_{ltU^{-1}}$-scale invariant. \ The $\sigma_{ltU^{-1}}$-scale
invariance of the two-field correlation function for zero-point radiation thus
requires
\begin{equation}
\left\langle \phi_{0}(ct,x)\phi_{0}(ct^{\prime},x^{\prime})\right\rangle
=\left\langle \phi_{0}(c\sigma t,\sigma x)\phi_{0}(c\sigma t^{\prime},\sigma
x^{\prime})\right\rangle
\end{equation}
From equation (26) in terms of the spectrum $f^{2}(|k|),$ this implies%
\begin{align}
&  \int_{-\infty}^{\infty}dkf^{2}(|k|)\cos[k(x-x^{\prime})-|k|c(t-t^{\prime
})]\nonumber\\
&  =\int_{-\infty}^{\infty}dkf^{2}(|k|)\cos[k(\sigma x-\sigma x^{\prime
})-|k|c(\sigma t-\sigma t^{\prime})]\nonumber\\
&  =\int_{-\infty}^{\infty}dk^{\prime}\left(  \sigma^{-1}f^{2}(|k^{\prime
}|/\sigma)\right)  \cos[k^{\prime}(x-x^{\prime})-|k^{\prime}|c(t-t^{\prime})]
\end{align}
where we have made the change of variable $k^{\prime}=\sigma k,$ $dk^{\prime
}=\sigma k^{\prime},$ $|k^{\prime}|=\sigma|k|,$ for $\sigma>0.$ \ Then the
correlation function is $\sigma_{ltU^{-1}}$-scale invariant provided $\left(
\sigma^{-1}f^{2}(|k^{\prime}|/\sigma)\right)  =f^{2}(|k^{\prime}|)$, which
means that
\begin{equation}
f^{2}(|k|)=const/|k|
\end{equation}
\ Thus we have found the $\sigma_{ltU^{-1}}$-scale-invariant spectrum in an
inertial frame in two spacetime dimensions. \ It turns out that this
$\sigma_{ltU^{-1}}$-scale invariant spectrum is also Lorentz invariant.

\subsubsection{Lorentz Invariance of Zero-Point Radiation}

Since the field $\phi$ is a Lorentz scalar, the Lorentz invariance of the
zero-point radiation corresponds to invariance of the correlation function
under the transformation $ct\rightarrow\gamma ct-\gamma\beta x,$
$x\rightarrow\gamma x-\gamma\beta ct;$ this corresponds to
\begin{equation}
\left\langle \phi_{0}(ct,x)\phi_{0}(ct^{\prime},x^{\prime})\right\rangle
=\left\langle \phi_{0}(\gamma ct-\gamma\beta x,\gamma x-\gamma\beta
ct)\phi_{0}(\gamma ct^{\prime}-\gamma\beta x^{\prime},\gamma x^{\prime}%
-\gamma\beta ct^{\prime})\right\rangle
\end{equation}
Now introducing the $\sigma_{ltU^{-1}}$-scale-invariant spectrum (30) into
integral (26) but using $k^{\prime}$ as the (dummy) variable of integration,
we have%
\begin{align}
&  <\phi_{0}(\gamma ct-\gamma\beta x,\gamma x-\gamma\beta ct)\phi_{0}(\gamma
ct^{\prime}-\gamma\beta x^{\prime},\gamma x^{\prime}-\gamma\beta ct^{\prime
})>=\nonumber\\
&  =\int_{-\infty}^{\infty}dk^{\prime}\frac{const}{|k^{\prime}|}\cos
[k^{\prime}\{\gamma(x-x^{\prime})-\gamma\beta c(t-t^{\prime})\}-|k^{\prime
}|\{\gamma c(t-t^{\prime})-\gamma\beta(x-x^{\prime})\}]\nonumber\\
&  =\int_{-\infty}^{\infty}dk^{\prime}\frac{const}{|k^{\prime}|}\cos[(\gamma
k^{\prime}+\gamma\beta|k^{\prime}|)(x-x^{\prime})-(\gamma|k^{\prime}%
|+\gamma\beta k^{\prime})c(t-t^{\prime})]\nonumber\\
&  =\int_{-\infty}^{\infty}dk\frac{const}{|k|}\cos[k(x-x^{\prime
})-|k|c(t-t^{\prime})]\nonumber\\
&  =<\phi_{0}(ct,x)\phi_{0}(ct^{\prime},x^{\prime})>
\end{align}
where we have changed the variable of integration to $k=\gamma k^{\prime
}+\gamma\beta|k^{\prime}|$ and have noted%
\begin{equation}
\frac{dk}{|k|}=\frac{\gamma(1+\beta k^{\prime}/|k^{\prime}|)dk^{\prime}%
}{\gamma|k^{\prime}|+\gamma\beta k^{\prime}}=\frac{dk^{\prime}}{|k^{\prime}|}%
\end{equation}
Thus indeed the integral takes exactly the same form in both the primed and
unprimed Lorentz frames, and the zero-point radiation spectrum is Lorentz
invariant, taking the same form in every inertial frame.

\subsubsection{Multiplicative Constant Determining Zero-Point Radiation}

The constant involved in the classical zero-point radiation spectrum can be
obtained from comparison with the experimentally observed Casimir forces
associated with random classical electromagnetic fields. \ It turns out that
this constant takes a numerical value which is immediately recognized as
corresponding to the familiar Planck constant appearing in the energy per
normal mode $U(k)=(1/2)\hbar c|k|=(1/2)\hbar\omega.$ \ By analogy with the
electromagnetic case, we choose the spectrum of random classical scalar
radiation to also give the same energy per normal mode so that%
\begin{equation}
\frac{1}{2}\hbar c|k|=U(k)=\frac{1}{8\pi}k^{2}f^{2}(|k|)
\end{equation}
The two-field correlation function\ then becomes
\begin{equation}
\left\langle \phi_{0}(ct,x)\phi_{0}(ct^{\prime},x^{\prime})\right\rangle
=\hbar c\int_{-\infty}^{\infty}\frac{dk}{|k|}\cos[k(x-x^{\prime}%
)-|k|c(t-t^{\prime})]
\end{equation}
Thus Planck's constant enters this classical analysis. \ In keeping with
nineteenth century ideas of physics, we could have expressed the unknown
spectral value $const=|k|f^{2}(|k|)$ in terms of Stefan's constant which was
measured in the 1870s; however, today Stefan's constant is far less familiar
than Planck's constant. \ The connection between the two is given by
$\sigma_{S}=(\pi^{2}k_{B}^{4})/(15\hbar^{3}c^{3}).$\cite{Morse339}

Although zero-point energy is familiar to students from its appearance in
quantum theory, most students are completely unaware of the possibility of
classical zero-point radiation. \ Also, most students are unaware of the fact
that an energy per normal mode given by a constant times the frequency of the
mode corresponds to a radiation spectrum which is both $\sigma_{ltU^{-1}}%
$-scale invariant and also Lorentz invariant. \ The spectrum of zero-point
radiation involves symmetries which are more fundamental than simply an
adjustment value for the ground state of the quantum harmonic oscillator.

\subsubsection{Closed-Form Evaluation of the Two-Field Correlation Function in
Zero-Point Radiation}

Since we know the exact form for the zero-point radiation spectrum, we can try
to evaluate the integral in Eq. (35) for the two-field correlation function.
\ Unfortunately the integrand is logarithmically divergent at small values of
the wave number $k$.\cite{infrared} \ Thus rather than evaluating the
expression in Eq. (35), we will follow Fulling and Davies\cite{F-D} and will
evaluate the expressions involving the derivatives $\left\langle \phi
_{0}(ct,x)\partial_{\mu^{\prime}}\phi_{0}(ct^{\prime},x^{\prime})\right\rangle
.$ \ Thus we have
\begin{equation}
\left\langle \phi_{0}(ct,x)\partial_{ct^{\prime}}\phi_{0}(ct^{\prime
},x^{\prime})\right\rangle =-\hbar c\int_{-\infty}^{\infty}\frac{dk}%
{|k|}|k|\sin[k(x-x^{\prime})-|k|c(t-t^{\prime})]
\end{equation}
and
\begin{equation}
\left\langle \phi_{0}(ct,x)\partial_{x^{\prime}}\phi_{0}(ct^{\prime}%
,x^{\prime})\right\rangle =\hbar c\int_{-\infty}^{\infty}\frac{dk}{|k|}%
k\sin[k(x-x^{\prime})-|k|c(t-t^{\prime})]
\end{equation}

Although the integrals in Eqs. (36) and (37) are now convergent at small wave
number $k,$ they still require a cut-off at large values of $|k|$ where the
trigonometric functions oscillate increasingly rapidly$.$ \ We will introduce
an exponential cut off of the integrand. \ Thus the sort of integral which we
must evaluate corresponds to the real or imaginary part of
\begin{align}
\lim_{\Lambda\rightarrow0_{+}}\int_{0}^{\infty}dkk^{n}\exp[(ib-\Lambda)k]  &
=\lim_{\Lambda\rightarrow0_{+}}\frac{\partial^{n}}{\partial(ib-\Lambda)^{n}%
}\int_{0}^{\infty}dk\exp[(ib-\Lambda)k]\nonumber\\
&  =\lim_{\Lambda\rightarrow0_{+}}\frac{\partial^{n}}{\partial(ib-\Lambda
)^{n}}\left[  \frac{\exp[(ib-\Lambda)k]}{(ib-\Lambda)}\right]  _{k=0}%
^{k=\infty}\nonumber\\
&  =\frac{(-1)^{n+1}(n)!}{(ib)^{n+1}}%
\end{align}
Thus we obtain the non-zero singular Fourier sine and cosine transforms%
\begin{equation}
\int_{0}^{\infty}dkk^{2m}\sin(bk)=\frac{(-1)^{2m}(2m)!}{b^{2m+1}}%
\end{equation}
and
\begin{equation}
\int_{0}^{\infty}dkk^{2m+1}\cos(bk)=\frac{(-1)^{2m+1}(2m+1)!}{b^{2m+2}}%
\end{equation}
Then it follows that for classical zero-point radiation, the correlation of
Eq. (37) becomes%
\begin{align}
\left\langle \phi_{0}(ct,x)\partial_{x^{\prime}}\phi_{0}(ct^{\prime}%
,x^{\prime})\right\rangle  &  =\hbar c\int_{-\infty}^{0}\frac{dk}{(-k)}%
k\sin[k(x-x^{\prime})-(-k)c(t-t^{\prime})]\nonumber\\
&  +\hbar c\int_{0}^{\infty}\frac{dk}{k}k\sin[k(x-x^{\prime})-kc(t-t^{\prime
})]\nonumber\\
&  =\hbar c\{\frac{1}{(x-x^{\prime})+c(t-t^{\prime})}+\frac{1}{(x-x^{\prime
})-c(t-t^{\prime})}\}\nonumber\\
&  =2\hbar c\frac{(x-x^{\prime})}{(x-x^{\prime})^{2}-c^{2}(t-t^{\prime})^{2}}%
\end{align}
and similarly Eq. (36) becomes%
\begin{align}
\left\langle \phi_{0}(ct,x)\partial_{ct^{\prime}}\phi_{0}(ct^{\prime
},x^{\prime})\right\rangle  &  =\hbar c\{\frac{1}{(x-x^{\prime})+c(t-t^{\prime
})}-\frac{1}{(x-x^{\prime})-c(t-t^{\prime})}\}\nonumber\\
&  =2\hbar c\frac{-c(t-t^{\prime})}{(x-x^{\prime})^{2}-c^{2}(t-t^{\prime}%
)^{2}}%
\end{align}
We notice that, with the extra minus sign between the two correlations, they
form a covariant (as opposed to contravariant) Lorentz vector, as would fit
with the covariant partial derivative. \ We also notice that there is no
parameter involving length, time, or energy which enters these expressions for
the classical zero-point field in an inertial frame.

\subsection{Thermal Radiation Spectrum Requirements in an Inertial Frame}

\subsubsection{$\sigma_{ltU^{-1}}$-Scaling Behavior and Wien's Law}

When discussing the spectrum of random classical radiation in an inertial
frame, we have so far considered only \textit{zero-point} radiation. \ The
spectrum of classical \textit{thermal} radiation must be connected
continuously with the spectrum of classical zero-point radiation as the
temperature goes to zero. \ Thus the \textit{thermal} spectrum $f^{2}(|k|,T)$
is a function of the wave number $k$ and the temperature $T$ such that
$f^{2}(|k|,T)\rightarrow4\pi\hbar c/|k|$ as $T\rightarrow0.$ \ We also know
that under an adiabatic change of the length of a box (or equivalently under a
$\sigma_{ltU^{-1}}$-scale change), the thermal radiation changes so that the
temperature $T\rightarrow T/\sigma.$ \ Thus for thermal radiation at non-zero
temperature, we go back to the two-point correlation function and require
$\sigma_{ltU^{-1}}$-scale\ invariance of the spectrum provided that the
temperature $T$ is rescaled to $T/\sigma,$ or $\left\langle \phi_{T/\sigma
}(c\sigma t,\sigma x)\phi_{T/\sigma}(c\sigma t^{\prime},\sigma x^{\prime
})\right\rangle =\left\langle \phi_{T}(ct,x)\phi_{T}(ct^{\prime},x^{\prime
})\right\rangle .$ \ The condition on the spectrum thus requires
\begin{align}
&  <\phi_{T/\sigma}(c\sigma t,\sigma x)\phi_{T/\sigma}(c\sigma t^{\prime
},\sigma x^{\prime})>=\nonumber\\
&  =\frac{1}{4\pi}\int_{-\infty}^{\infty}dkf^{2}(|k|,T/\sigma)\cos[k(\sigma
x-\sigma x^{\prime})-|k|c(\sigma t-\sigma t^{\prime})]\nonumber\\
&  =\frac{1}{4\pi}\int_{-\infty}^{\infty}dkf^{2}(|k|,T/\sigma)\cos[\sigma
k(x-x^{\prime})-|\sigma k|c(t-t^{\prime})]\nonumber\\
&  =\frac{1}{4\pi}\int_{-\infty}^{\infty}(dk^{\prime}/\sigma)f^{2}(|k^{\prime
}/\sigma|,T/\sigma)\cos[k^{\prime}(x-x^{\prime})-|k^{\prime}|c(t-t^{\prime
})]\nonumber\\
&  =\frac{1}{4\pi}\int_{-\infty}^{\infty}dk^{\prime}f^{2}(|k^{\prime}%
|,T)\cos[k^{\prime}(x-x^{\prime})-|k^{\prime}|c(t-t^{\prime})]\nonumber\\
&  =<\phi_{T}(ct,x)\phi_{T}(ct^{\prime},x^{\prime})>
\end{align}
Thus the required condition on the spectrum corresponds to
\begin{equation}
\frac{1}{\sigma}f^{2}(|k|/\sigma,T/\sigma)=f^{2}(|k|,T)
\end{equation}
for all positive values of $\sigma$. \ The condition (44) has the general
solution%
\begin{equation}
f^{2}(|k|,T)=\frac{1}{|k|}g\left(  \frac{|k|}{T}\right)
\end{equation}
for an arbitrary function $g(|k|/T).$ \ This result corresponds to Wien's
displacement law for the thermal radiation spectrum.\cite{Lavenda67-70} \ Thus
in an inertial frame, Wien's displacement law follows from the $\sigma
_{ltU^{-1}}$-scaling requirements of the theory.

In the limit as the temperature goes to zero, we must recover the zero-point
radiation spectrum. \ Also, we define the scale of temperature such that the
energy of a normal mode in Eq. (21) is given by $k_{B}T$ at large temperature.
\ Thus the limits on the function $g$ require from
\begin{equation}
U(k)=(1/8\pi)k^{2}f^{2}(|k|,T)=(1/8\pi)|k|g^{2}(|k|/T)
\end{equation}
that%
\begin{equation}
g\left(  \frac{|k|}{T}\right)  \rightarrow4\pi\hbar c\ \ for\ k_{B}T<<\hbar
c|k|\text{ \ \ and}\ g\left(  \frac{|k|}{T}\right)  \rightarrow8\pi\frac
{k_{B}T}{|k|}\ for\ T>>\hbar c|k|
\end{equation}
Also the function $g(|k|/T)$ must be a monotonically decreasing function of
its argument so that the thermal radiation spectrum will be a monotonically
increasing function of temperature. Thus determination of the spectrum of
classical thermal radiation involves determining a single unknown function
$g(|k|/T),$ and all thermal radiation spectra involve rescaled versions of
this function$.$ \ If we knew the spectrum for even one temperature $T>0$,
then we could determine the spectrum for any other temperature $T^{\prime
}=\sigma T$ by simply carrying out a $\sigma_{ltU^{-1}}$-scale change on the
wave number $k,$ corresponding to%
\begin{align}
f^{2}(|k|,T^{\prime})  &  =f^{2}(|k|,\sigma T)=(1/|k|)g(|k|/\sigma
T)=\nonumber\\
(1/\sigma)/(|k|/\sigma)g((|k|/\sigma)/T)  &  =(1/\sigma)f^{2}(|k|/\sigma,T).
\end{align}

\subsubsection{Consistency with the Stefan-Boltzmann Relation}

From the functional form for thermal radiation corresponding to (45), it
immediately follows that the thermal energy density $u_{T}$ follows the
Stefan-Boltzmann relation obtained earlier from pure thermodynamics in Eq.
(14). \ Thus the thermal energy density $u_{T}$ can be evaluated as the finite
energy density above the divergent zero-point energy%
\begin{align}
u_{T}(x)  &  =\frac{1}{8\pi}\left[  \frac{1}{c^{2}}\left(  \frac{\partial
\phi_{T}}{\partial t}\right)  ^{2}+\left(  \frac{\partial\phi_{T}}{\partial
x}\right)  ^{2}\right]  -\frac{1}{8\pi}\left[  \frac{1}{c^{2}}\left(
\frac{\partial\phi_{0}}{\partial t}\right)  ^{2}+\left(  \frac{\partial
\phi_{0}}{\partial x}\right)  ^{2}\right] \nonumber\\
&  =\frac{1}{8\pi}\left[  \left(  \frac{1}{c^{2}}\frac{\partial}{\partial
t}\frac{\partial}{\partial t^{\prime}}+\frac{\partial}{\partial x}%
\frac{\partial}{\partial x^{\prime}}\right)  \left\{  <\phi_{T}(ct,x)\phi
_{T}(ct^{\prime},x^{\prime})>-<\phi_{0}(ct,x)\phi_{0}(ct^{\prime},x^{\prime
})>\right\}  \right]  _{t=t^{\prime},x=x^{\prime}}\nonumber\\
&  =\frac{1}{8\pi}\left[  \frac{1}{4\pi}\int_{-\infty}^{\infty}dk\frac{1}%
{|k|}[g\left(  |v|\right)  -4\pi\hbar c]2k^{2}\cos[k^{\prime}(x-x^{\prime
})-|k^{\prime}|c(t-t^{\prime})]\right]  _{t=t^{\prime},x=x^{\prime}%
}\nonumber\\
&  =\left(  \frac{1}{4\pi}\right)  ^{2}\int_{-\infty}^{\infty}dk|k|[g\left(
|v|\right)  -4\pi\hbar c]=T^{2}\left[  \left(  \frac{1}{4\pi}\right)  ^{2}%
\int_{-\infty}^{\infty}dv|v|[g\left(  |v|\right)  -4\pi\hbar c]\right]
=\sigma_{2D}T^{2}%
\end{align}
where we have subtracted the integrands at given wave number $k,$\ and we have
used the substitution $v=k/T.$ \ The final integral gives the constant
$\sigma_{2D}$\ which is independent of temperature $T,$ and the result
corresponds exactly to the thermodynamic result in Eq. (14).

\subsubsection{Correlation Lengths and Times for Thermal Radiation in an
Inertial Frame}

Thermal radiation at non-zero temperature in a finite-length box involves a
finite amount of radiation energy which is distributed over the radiation
normal modes of the box. \ The thermal radiation is energy above the
$\sigma_{ltU^{-1}}$-scale-invariant zero-point radiation. \ Since there are an
infinite number of normal modes of ever-increasing frequency, the finite
amount of thermal radiation must decrease as the frequency of the modes
increases. \ We expect thermal radiation to be distributed in a smooth
monotonic fashion with more thermal energy at lower frequencies. \ Since the
zero-point energy per normal mode increases with frequency, there must be some
normal mode where the thermal energy in the mode is comparable to the
zero-point energy in the mode. \ The wavelength $\lambda_{T}$, frequency
$\nu_{T}$, and energy $U(\nu_{T},T)$ of this normal mode will provide a
characteristic length, time, and energy related to the temperature for the
thermal radiation distribution; thus we expect from Wien's displacement
theorem $\lambda_{T}=const/T,$ $\nu_{T}=c/\lambda_{T}=cT/const,$ and
$U(\nu_{T},T)=const^{\prime}\times T$ . \ Since all three parameters scale
together in a thermal distribution, knowledge of any one parameter implies
information about the other two. \ These parameters will be directly related
to correlation lengths and correlation times for the thermal radiation. \ We
emphasize that the $\sigma_{ltU^{-1}}$-scale-invariant zero-point radiation
has no such parameters in an inertial frame. \ Thus in an inertial frame,
thermal radiation at non-zero-temperature $T$ is crucially different from
zero-point radiation.

\subsubsection{Correlation Function for a Single Spatial Point or Single Time
for Thermal Radiation in an Inertial Frame}

From the scaling information above in Eqs. (43) and (45), the two-point
correlation function for the thermal radiation fields takes the form%
\begin{equation}
<\phi_{T}(ct,x)\phi_{T}(ct^{\prime},x^{\prime})>=\frac{1}{4\pi}\int_{-\infty
}^{\infty}\frac{dk}{|k|}g(|k|/T)\cos[k(x-x^{\prime})-|k|c(t-t^{\prime})]
\end{equation}
We would like to turn our information about the spectrum $g(|k|/T)$ into
information about the time correlations at a single spatial point for use in
our later analysis in a Rindler frame. \ In order to assure convergence of the
integral at small values of $k$, we again consider the time derivative of Eq.
(49) at $x=x^{\prime},$%
\begin{align}
&  <\phi_{T}(ct,x)\partial_{ct^{\prime}}\phi_{T}(ct^{\prime},x)>=\frac{1}%
{4\pi}\int_{-\infty}^{\infty}dkg(|k|/T)\sin[|k|c(t-t^{\prime})]\nonumber\\
&  =T\frac{1}{4\pi}\int_{0}^{\infty}dv[2g(|v|)]\sin[|v|Tc(t-t^{\prime
})]=TF[Tc(t-t^{\prime})]
\end{align}
where we have introduced the change of integration variable $v=k/T$ for
non-zero temperature $T.$ \ Here the expression $F[Tc(t-t^{\prime})]$ is some
unknown function of the temperature $T$\ times the time difference
$c(t-t^{\prime})$. \ Because a rapidly varying sine function of
$|k|Tc(t-t^{\prime})$ should cause cancellations between the positive and
negative contributions in the integrand of Eq. (50), we would expect that at
small time differences the correlation function should reflect the zero-point
radiation spectrum $g(|k|/T)\rightarrow4\pi\hbar c$ at high frequencies, while
at large time differences, the correlation function should reflect the
low-frequency thermal radiation spectrum $g(|k|/T)\rightarrow8\pi k_{B}T/|k|$
as in Eq. (47). \ Also, at zero temperature, the correlation function should
depend solely on the scale-invariant zero-point spectrum and therefore should
depend upon the time difference but not on any parameter involving length,
time, or energy. \ Thus we expect the limits
\begin{equation}
TF[Tc(t-t^{\prime})]\rightarrow2\hbar c\frac{1}{c(t-t^{\prime})}\text{
\ \ \ for }Tc(t-t^{\prime})<<1
\end{equation}
in connection with Eq. (42) for zero-point radiation, and we expect the limit
\begin{equation}
TF[Tc(t-t^{\prime})]\rightarrow const\times T\ \ \ \text{ for }Tc(t-t^{\prime
})>>1
\end{equation}
in connection with the Stefan-Boltzmann relation (14) and the energy density
(49). \ Thus thermal radiation is a one-parameter family of radiation spectra
determined by the spectral function $g(|k|/T)$ or by the correlation function
$F[Tc(t-t^{\prime})].$

We can also consider the correlation function at a fixed time $t=t^{\prime}$
at two different spatial points $x$ and $x^{\prime}.$ \ The correlation with a
non-zero derivative corresponds to the spatial derivative, which from Eq. (50)
becomes%
\begin{align}
&  <\phi_{T}(ct,x)\partial_{x^{\prime}}\phi_{T}(ct^{\prime},x)>_{T}=\frac
{1}{4\pi}\int_{-\infty}^{\infty}\frac{dk}{|k|}g(|k|/T)k\sin[k(x-x^{\prime
})]\nonumber\\
&  =T\frac{1}{4\pi}\int_{0}^{\infty}dv[2g(|v|)]\sin[|v|T(x-x^{\prime
})]=TF[T(x-x^{\prime})]
\end{align}
We see that exactly the same unknown function $F$ is involved in the
correlation function at a single time as is involved at a single spatial
point. \ Thus in an inertial frame, we obtain the same information from the
time correlations as from the spatial correlations of the scalar radiation field.

\section{Part II: Thermal Radiation in a Rindler Frame}

\subsection{Introduction of a Rindler Frame}

\subsubsection{Analogy with Boltzmann's Use of Gravity}

Classical thermal radiation in an inertial frame seems to give no hint
regarding the form of the spectrum at non-zero-temperature. \ In an inertial
frame, zero-point radiation is $\sigma_{ltU^{-1}}$-scale invariant, and the
correlation functions in Eqs. (41)\ and (42) simply scale as $\sigma^{-1}$
without giving any information about a functional form at
non-zero-temperature. \ In order to get further information about the thermal
spectrum, we do the same thing which Boltzmann did when dealing with the
thermal velocity distribution for free particles in a box; he introduced a
gravitational field and assumed that basic ideas of thermal equilibrium
applied also in a gravitational field. \ Now by the equivalence principle, a
gravitational field is locally equivalent to a coordinate frame undergoing
constant acceleration relative to an inertial frame. \ Thus we will consider a
one-dimensional box undergoing uniform acceleration relative to an inertial
frame, and will assume that the basic ideas of thermodynamics hold in this
accelerating frame. \ 

\subsubsection{Rindler Coordinates}

Although the nonrelativistic mechanics of Boltzmann's analysis\cite{BoltzMax}
allows a single constant acceleration throughout an accelerating box, this is
not true in a relativistic analysis. \ The closest that we can come to a
constant gravitational field is that provided by a Rindler coordinate
system\cite{Schutz}\cite{Rindler} which is accelerating relative to an
inertial frame with which it is instantaneously at rest at time $t=0.$ \ If
the time and space coordinates of the inertial frame are given by $(ct,x),$
then the connections with the time and space coordinates $(\eta,\xi)\,\ $of
the Rindler frame are%
\begin{equation}
ct=\xi\sinh\eta
\end{equation}%
\begin{equation}
x=\xi\cosh\eta
\end{equation}
with $-\infty<\eta<\infty,$ and $0<\xi$. \ Using the relationship $\cosh
^{2}\eta-\sinh^{2}\eta=1,$ it follows that a point with fixed spatial
coordinate $\xi$ follows a trajectory%
\begin{equation}
x=(\xi^{2}+c^{2}t^{2})^{1/2}%
\end{equation}
in the inertial frame, and therefore undergoes a constant proper acceleration
given by
\begin{equation}
a_{\xi}=\left(  \frac{d^{2}x}{dt^{2}}\right)  _{t=0}=\frac{c^{2}}{\xi}%
\end{equation}
Thus the proper acceleration $a_{\xi}$ of a coordinate point with fixed
spatial coordinate $\xi$ becomes smaller as $\xi$ increases and becomes larger
for small $\xi.$ \ As $\xi\rightarrow0,$ the acceleration diverges, and the
point where $\xi=0$ corresponds to an event horizon. \ 

Now in thermal equilibrium in a gravitational field treated within a
relativistic theory, the temperature must increase at points which are lower
and must decrease at points which are higher according to the Tolman-Ehrenfest
relation\cite{T-E} $T(g_{00})^{1/2}=const.$ \ Here in our two-dimensional
spacetime, we have from Eqs. (1), (55), and (56)
\begin{equation}
ds^{2}=c^{2}dt^{2}-dx^{2}=\xi^{2}d\eta^{2}-d\xi^{2}%
\end{equation}
so that $g_{00}=\xi^{2},$ and therefore in the Rindler frame%
\begin{equation}
T\xi=const
\end{equation}
Thus, except at absolute zero, there is no single temperature which can be
assigned to a box in a Rindler frame; rather, for $T>0,$ the temperature at
the bottom of the box must be larger than at the top of the box. \ Again under
adiabatic compression, we expect the temperature in the box to increase. \ The
temperature within a Rindler frame must be characterized by the $\sigma
_{ltU^{-1}}$-scale-invariant constant which enters the Tolman-Ehrenfest
relation (60). \ 

\subsection{Radiation Normal Modes in a Rindler Frame}

At this point, we wish to consider the spectrum of random radiation as seen in
the Rindler frame. \ First we obtain the radiation normal modes. \ The wave
equation (3) in an inertial frame can be transformed to the wave equation in
the Rindler frame by using the transformation equations (55), (56), together
with the scalar behavior of the field $\phi$ under a coordinate
transformation. \ The scalar field takes the same value in any coordinate
frame. \ Thus the field $\varphi(\eta,\xi)$ in the Rindler frame is equal to
the field $\phi(ct,x)$ in the inertial frame at the same spacetime point,%
\begin{equation}
\varphi(\eta,\xi)=\phi(ct,x)=\phi(\xi\sinh\eta,\xi\cosh\eta)
\end{equation}
\ Then using the usual rules for partial derivatives, we find that the wave
equation (3) becomes in the Rindler frame%
\begin{equation}
\left(  \frac{\partial^{2}\varphi}{\partial\xi^{2}}\right)  +\frac{1}{\xi
}\left(  \frac{\partial\varphi}{\partial\xi}\right)  -\frac{1}{\xi^{2}}\left(
\frac{\partial^{2}\varphi}{\partial\eta^{2}}\right)  =0
\end{equation}
The solutions of this Rindler wave equation take the form $H(\ln\xi\pm\eta)$
where $H$ is an arbitrary function.\ \ Thus whereas the general solution of
the scalar wave equation (3) in an inertial frame is $\phi(ct,x)=h_{+}%
(x-ct)+h_{-}(x+ct)$ where $h_{+}$ and $h_{-}$ are arbitrary functions, the
general solution in a Rindler frame is $\varphi(\eta,\xi)=H_{+}(\ln\xi
-\eta)+H_{-}(\ln\xi+\eta)$ where $H_{+}$ and $H_{-}$ are arbitrary
functions$.$ \ The normal mode solutions of the wave equation in the Rindler
frame for a box extending from $0<\xi=a$ to $\xi=b$ with Dirichlet boundary
conditions can be obtained by separation of variables and can be written as
\begin{equation}
\varphi_{n}(\eta,\xi)=\mathcal{F}_{n}\left(  \frac{2}{\ln(b/a)}\right)
^{1/2}\sin\left[  \frac{n\pi}{\ln(b/a)}\ln\left(  \frac{\xi}{a}\right)
\right]  \cos\left[  \frac{n\pi}{\ln(b/a)}\eta+\theta_{n}\right]  ,\text{
\ \ }n=1,2,3...
\end{equation}
where the spatial functions
\begin{equation}
\psi_{n}(\eta,\xi)=\left(  \frac{2}{\ln(b/a)}\right)  ^{1/2}\sin\left[
\frac{n\pi}{\ln(b/a)}\ln\left(  \frac{\xi}{a}\right)  \right]
\end{equation}
arise from a Sturm-Liouville system\cite{MW} and so form a complete
orthonormal set with weight $1/\xi$ on the interval $a<\xi<b$%
\begin{align}
\int_{a}^{b}\frac{d\xi}{\xi}\psi_{n}(\xi)\psi_{m}(\xi)  &  =\nonumber\\
&  =\int_{a}^{b}\frac{d\xi}{\xi}\frac{2}{\ln(b/a)}\sin\left[  \frac{n\pi}%
{\ln(b/a)}\ln\left(  \frac{\xi}{a}\right)  \right]  \sin\left[  \frac{m\pi
}{\ln(b/a)}\ln\left(  \frac{\xi}{a}\right)  \right] \nonumber\\
&  =\int_{v=0}^{v=\pi}\frac{\ln(b/a)}{\pi}dv\frac{2}{\ln(b/a)}\sin nv\sin
mv\nonumber\\
&  =\delta_{nm}%
\end{align}
where we have used the substitution $v=[\pi\ln(\xi/a)]/\ln(b/a)~$in evaluating
the integral. \ For a radiation normal mode, the Rindler time parameter $\eta$
agrees with all local clocks when adjusted by $\xi;$ thus the time $t=\xi\eta$
gives the proper time of a clock located at fixed Rindler spatial coordinate
$\xi.$ \ 

For time-stationary random radiation, the field $\varphi(\eta,\xi)$ can be
written as a sum over the normal modes $\varphi_{n}(\eta,\xi)$ in (62) with
random phases $\theta_{n}$ distributed randomly over the interval $[0,2\pi)$
and distributed independently for each value of $n.$ \ Then the two-field
correlation function is obtained in analogy with Eq. (22)-(24). \ For a large
box, we can go through the same sort of analysis as is given in Eq. (24)-(26)
to obtain a set of normal modes dependent upon a continuous wave number
$\kappa=n\pi/\ln(b/a)$ and can accommodate both cosine functions by including
the integral over negative values of $\kappa,$ exactly as in Eq. (26). \ Here
for the Rindler frame, we obtain the expression analogous to Eq. (26) in an
inertial frame,%
\begin{equation}
\left\langle \varphi(\eta,\xi)\varphi(\eta^{\prime},\xi^{\prime})\right\rangle
=\frac{1}{4\pi}\int_{-\infty}^{\infty}d\kappa\mathcal{F}^{2}(|\kappa
|)\cos[\kappa(\ln\xi-\ln\xi^{\prime})-|\kappa|(\eta-\eta^{\prime})]
\end{equation}

\subsection{$\sigma_{ltU^{-1}}$-Scale Change in a Rindler Frame}

Now we would like to consider a $\sigma_{ltU^{-1}}$-scale transformation in a
Rindler frame. \ Earlier we found that $\sigma_{ltU^{-1}}$-scale
invariance~was a strong condition in an inertial frame. \ However, for a
Rindler frame, we notice from the coordinate transformation equations
$ct=\xi\sinh\eta$ and $x=\xi\cosh\eta$ that a $\sigma_{ltU^{-1}}$-scale change
$x\rightarrow x^{\prime}=\sigma x,$ $t\rightarrow t^{\prime}=\sigma t$ will
affect the Rindler spatial coordinate $\xi$ but not the Rindler time
coordinate $\eta;$ thus we have $\xi\rightarrow\xi^{\prime}=\sigma\xi,$ but
$\eta\rightarrow\eta^{\prime}=\eta,$ and $ds^{2}\rightarrow ds^{\prime
2}=\sigma^{2}ds^{2}=\xi^{\prime2}d\eta^{\prime2}-d\xi^{\prime2}.$ It follows
from Eq. (65) and $\ln(\sigma\xi)-\ln(\sigma\xi^{\prime})=\ln\xi-\ln
\xi^{\prime}$ that under a $\sigma_{ltU^{-1}}$-scale change, we have%
\begin{align}
\left\langle \varphi(\eta,\sigma\xi)\varphi(\eta^{\prime},\sigma\xi^{\prime
})\right\rangle  &  =\frac{1}{4\pi}\int_{-\infty}^{\infty}d\kappa
\mathcal{F}^{2}(|\kappa|)\cos[\kappa\{\ln(\xi)-\ln(\xi^{\prime})\}-|\kappa
|(\eta-\eta^{\prime})]\nonumber\\
&  =\left\langle \varphi(\eta,\xi)\varphi(\eta^{\prime},\xi^{\prime
})\right\rangle
\end{align}
But then $\sigma_{ltU^{-1}}$-scale invariance gives no condition whatsoever on
the spectrum. Any spectrum which is time-stationary in a Rindler frame is
$\sigma_{ltU^{-1}}$-scale invariant. \ The scaling operation simply carries
the spatial points to new locations within the Rindler frame without imposing
any restriction on the time-stationary spectrum of random radiation. \ This
result is completely different from the requirements of $\sigma_{ltU^{-1}}%
$-scale invariance in an inertial frame.

The $\sigma_{ltU^{-1}}$-scale invariance of thermal radiation in a Rindler
frame is a reminder that the temperature of thermal radiation in a large box
in a Rindler frame has no unique value but rather changes continuously
throughout the box. \ Thermal radiation in a Rindler frame is characterized by
a $\sigma_{ltU^{-1}}$-scale-invariant constant which is related to the
Tolman-Ehrenfest condition $T\xi=const.$ \ Under a $\sigma_{ltU^{-1}}$-scale
transformation, the temperature $T$ at a spatial point $\xi$ is mapped so that
$T\rightarrow T^{\prime}=T/\sigma$ while the spatial coordinate $\xi$ is
mapped so that $\xi\rightarrow\xi^{\prime}=\sigma\xi,$ leaving unchanged the
relationship $T^{\prime}\xi^{\prime}=(T/\sigma)(\sigma\xi)=const$

\subsection{Zero-Point Radiation in a Rindler Frame}

Classical physics deals with tensor quantities, and the zero-point radiation
which is present in an inertial frame must also be present in a Rindler frame
which is accelerating relative to an inertial frame. \ Furthermore, we expect
that the thermal radiation which is present in a Rindler frame must fit with
the zero-point radiation in the Rindler frame. \ We can obtain the form taken
by the random zero-point radiation in the Rindler frame by noting that the
field $\phi$ is a scalar under coordinate transformation so that the field
$\varphi(\eta,\xi)$ in the Rindler frame is equal to the field $\phi(ct,x)$ in
the inertial frame at the same spacetime point, as given in Eq. (61).
\ Therefore we have from Eqs. (41) and (42) and from the transformations of
Eqs. (55)\ and (56)
\begin{align}
&  <\varphi_{0}(\eta,\xi)\frac{\partial}{\partial\eta^{\prime}}\varphi
_{0}(\eta^{\prime},\xi^{\prime})>=\nonumber\\
&  =<\phi(ct,x)\frac{\partial}{\partial ct^{\prime}}\phi(ct^{\prime}%
,x^{\prime})>\frac{\partial ct^{\prime}}{\partial\eta^{\prime}}+<\phi
(ct,x)\frac{\partial}{\partial x^{\prime}}\phi(ct^{\prime},x^{\prime}%
)>\frac{\partial x^{\prime}}{\partial\eta^{\prime}}\nonumber\\
&  =2\hbar c\frac{-c(t-t^{\prime})}{(x-x^{\prime})^{2}-c^{2}(t-t^{\prime}%
)^{2}}\xi^{\prime}\cosh\eta^{\prime}+2\hbar c\frac{(x-x^{\prime}%
)}{(x-x^{\prime})^{2}-c^{2}(t-t^{\prime})^{2}}\xi^{\prime}\sinh\eta^{\prime
}\nonumber\\
&  =2\hbar c\frac{\xi\xi^{\prime}\sinh(\eta-\eta^{\prime})}{\xi^{2}%
+\xi^{\prime2}-2\xi\xi^{\prime}\cosh(\eta-\eta)}%
\end{align}
and similarly%
\begin{align}
&  <\varphi_{0}(\eta,\xi)\frac{\partial}{\partial\xi^{\prime}}\varphi_{0}%
(\eta^{\prime},\xi^{\prime})>=\nonumber\\
&  =<\phi(ct,x)\frac{\partial}{\partial ct^{\prime}}\phi(ct^{\prime}%
,x^{\prime})>\frac{\partial ct^{\prime}}{\partial\xi^{\prime}}+<\phi
(ct,x)\frac{\partial}{\partial x^{\prime}}\phi(ct^{\prime},x^{\prime}%
)>\frac{\partial x^{\prime}}{\partial\xi^{\prime}}\nonumber\\
&  =2\hbar c\frac{-c(t-t^{\prime})}{(x-x^{\prime})^{2}-c^{2}(t-t^{\prime}%
)^{2}}\sinh\eta^{\prime}+2\hbar c\frac{(x-x^{\prime})}{(x-x^{\prime}%
)^{2}-c^{2}(t-t^{\prime})^{2}}\cosh\eta^{\prime}\nonumber\\
&  =2\hbar c\frac{\xi\cosh(\eta-\eta^{\prime})-\xi^{\prime}}{\xi^{2}%
+\xi^{\prime2}-2\xi\xi^{\prime}\cosh(\eta-\eta)}%
\end{align}
We notice immediately that these expressions involve only the time difference
$\eta-\eta^{\prime}$ in the Rindler frame, and not the specific times $\eta$
and $\eta^{\prime}$. \ Thus the zero-point radiation is a time-stationary
distribution in a Rindler frame. \ Indeed the $\sigma_{ltU^{-1}}%
$-scale-invariant and Lorentz-invariant zero-point radiation is the only
spectrum of random classical radiation which is time-stationary in both all
inertial frames and all Rindler frames. \ 

At a single time $\eta=\eta^{\prime},$ but at two different spatial points
$\xi$ and $\xi^{\prime},$ we find~from Eqs. (68) and (69) that $\left\langle
\varphi_{0}(\eta,\xi)\frac{\partial}{\partial\eta^{\prime}}\varphi_{0}%
(\eta,\xi^{\prime})\right\rangle =0,$ while%
\begin{equation}
\left\langle \varphi_{0}(\eta,\xi)\frac{\partial}{\partial\xi^{\prime}}%
\varphi_{0}(\eta,\xi^{\prime})\right\rangle =\frac{2\hbar c}{\xi-\xi^{\prime}}%
\end{equation}
However, this equation (70) is exactly the same correlation function as
appears in Eq. (41) for zero-point radiation in an inertial frame when
$t=t^{\prime}$. \ Indeed, a single time in a Rindler frame is also a single
time in the inertial frame with which the Rindler frame is momentarily at
rest, so that the correlation function involving different spatial points
given in Eq. (70) can be interpreted as the correlation function at a single
time in either the Rindler frame or the associated inertial frame. \ Thus this
correlation function in the Rindler frame gives us no new information beyond
that of an inertial frame.

At a single Rindler spatial coordinate $\xi,$ but at two different times
$\eta$ and $\eta^{\prime},$ the correlation functions (68) and (69) become%
\begin{align}
\left\langle \varphi_{0}(\eta,\xi)\frac{\partial}{\partial\eta^{\prime}%
}\varphi_{0}(\eta^{\prime},\xi^{\prime})\right\rangle _{\xi^{\prime}=\xi}  &
=2\hbar c\frac{2\sinh[(\eta-\eta^{\prime})/2]\cosh[(\eta-\eta^{\prime}%
)/2]}{4\sinh^{2}[(\eta-\eta^{\prime})/2]}\nonumber\\
&  =\hbar c\coth\left(  \frac{\eta-\eta^{\prime}}{2}\right)
\end{align}
and%
\begin{equation}
\left\langle \varphi_{0}(\eta,\xi)\frac{\partial}{\partial\xi^{\prime}}%
\varphi_{0}(\eta^{\prime},\xi^{\prime})\right\rangle _{\xi^{\prime}=\xi
}=-\frac{\hbar c}{\xi}%
\end{equation}
These expressions are quite different from the analogous expressions for
zero-point radiation in an inertial frame found from (41) and (42) when
$x=x^{\prime}$.

\subsection{From Zero-Point Radiation to Thermal Radiation in a Rindler Frame}

\subsubsection{Scaling Parameter for Thermal Radiation in a Rindler Frame}

We notice that the \textit{zero-point }expression $\left\langle \varphi
_{0}(\eta,\xi)\partial_{\eta^{\prime}}\varphi_{0}(\eta^{\prime},\xi^{\prime
})\right\rangle _{\xi^{\prime}=\xi}$ given in Eq. (71) has no dependence upon
the spatial coordinate $\xi,$ but does have a dependence upon the Rindler time
difference $\eta-\eta^{\prime}.$ \ We expect the \textit{thermal} correlation
function $<\varphi_{\alpha}(\eta,\xi)\partial_{\eta^{\prime}}\varphi_{\alpha
}(\eta^{\prime},\xi^{\prime})>_{\xi^{\prime}=\xi}$ in a Rindler frame to
depend upon a single $\sigma_{ltU^{-1}}$-scale-invariant parameter $\alpha$
which can take on real values corresponding to the relationship $T\xi=const.$
\ The time correlation at a single spatial coordinate at non-zero temperature,
when expressed in terms of the Rindler time parameter $\eta,$ should be the
same at any point of the Rindler frame with no dependence upon the position
coordinate $\xi.$ \ Furthermore, the time correlation function must involve
the thermal parameter $\alpha$ as a factor in connection with the time
difference $\eta-\eta^{\prime}$. \ From our considerations of the time
correlation for thermal radiation in an inertial frame appearing in connection
with Eq. (51), it is natural to insert a $\sigma_{ltU^{-1}}$-scale-invariant
parameter $\alpha$ into the Rindler time correlation (71) as
\begin{equation}
\left\langle \varphi_{\alpha}(\eta,\xi)\frac{\partial}{\partial\eta^{\prime}%
}\varphi_{\alpha}(\eta^{\prime},\xi^{\prime})\right\rangle _{\xi^{\prime}=\xi
}=\alpha\hbar c\coth\left(  \alpha\frac{\eta-\eta^{\prime}}{2}\right)
\end{equation}
When $\alpha=1,$ the situation corresponds to zero-point radiation in the
Rindler frame, as given in Eq. (71). \ When $\alpha>1,$ the situation
corresponds to thermal radiation at a temperature greater than absolute zero.
\ We notice that the thermal correlation given in Eq. (73) is a monotonically
increasing function of $\alpha$ which has the asymptotic limits required for
thermal radiation in Eqs. (52) and (53). \ 

\subsubsection{Thermal Radiation Spectrum in a Rindler Frame}

The spectrum $\mathcal{F}_{\alpha}^{2}(|\kappa|)$ for thermal radiation in a
Rindler frame can be obtained from the correlation-function expressions given
in Eqs. (66) and (73) which require%
\begin{align}
\left\langle \varphi_{\alpha}(\eta,\xi)\frac{\partial}{\partial\eta^{\prime}%
}\varphi_{\alpha}(\eta^{\prime},\xi^{\prime})\right\rangle _{\xi^{\prime}%
=\xi}  &  =\frac{1}{4\pi}\int_{-\infty}^{\infty}d\kappa\mathcal{F}_{\alpha
}^{2}(|\kappa|)|\kappa|\sin[|\kappa|(\eta-\eta^{\prime})]\nonumber\\
&  =\alpha\hbar c\coth\left(  \alpha\frac{\eta-\eta^{\prime}}{2}\right)
\end{align}
We can take the inverse Fourier sine transform of Eq. (74). \ The singular
Fourier sine transform can be broken into a singular piece involving the same
integral as in Eq. (39) and also a convergent piece which can be found in a
standard table of integrals.\cite{GR} \ Thus we have%
\begin{align}
\mathcal{F}_{\alpha}^{2}(|\kappa|)|\kappa|  &  =4\int_{0}^{\infty}d\eta\left[
\alpha\hbar c\coth\left(  \alpha\frac{\eta}{2}\right)  \right]  \sin
(|\kappa|\eta)\nonumber\\
&  =4\int_{0}^{\infty}d\eta(\alpha\hbar c)\left[  1+\frac{2}{\exp(\alpha
\eta)-1}\right]  \sin(|\kappa|\eta)\nonumber\\
&  =4(\hbar c)\left[  \frac{\alpha}{|\kappa|}+2\left\{  \frac{\pi}{2}%
\coth\left(  \frac{|\kappa|\pi}{\alpha}\right)  -\frac{\alpha}{2|\kappa
|}\right\}  \right] \nonumber\\
&  =4\pi\hbar c\coth\left(  \frac{|\kappa|\pi}{\alpha}\right)
\end{align}
as the spectrum for thermal radiation in a Rindler frame. \ 

\subsubsection{The Vacuum State in a Rindler Frame}

In equation (75), the normal-mode spectrum for non-zero thermal radiation
involves $\alpha>1,$ whereas the spectrum corresponding to zero-point
radiation corresponds to $\alpha=1.$ \ Thus zero-point radiation as seen in a
Rindler frame has the same basic spectral form as does thermal radiation.
\ This situation is quite different from that of an inertial frame where
zero-point radiation involves the power-law spectrum of Eq. (34)
$f^{2}(|k|)=4\pi\hbar c/|k|$ and gives no hint of the spectral form for
non-zero temperature. \ It seems to come as a shock that in a Rindler frame
the vacuum state does not have a spectrum of random radiation involving a
power of the Rindler frequency
%TCIMACRO{\TEXTsymbol{\vert}}%
%BeginExpansion
$\vert$%
%EndExpansion
$\kappa|$ but rather has a more complicated functional form reflecting the
non-inertial character of the coordinate frame. \ An analogous situation
arises for the electric field lines of a point charge; the electric field
lines point radially away from the charge in all inertial frames, and yet
change character so as to "droop" when undergoing the acceleration of a
Rindler frame.

\subsection{Transferring Results from a Rindler Frame to an Inertial Frame}

Once we have the spectrum for classical thermal radiation in a Rindler frame,
we can transfer this information back to an inertial frame by treating a box
of thermal radiation which is far from the event horizon of the Rindler frame.
\ The acceleration of a fixed spatial coordinate $\xi$ is given by $a_{\xi
}=c^{2}/\xi$ and so becomes smaller as the distance $\xi$ from the event
horizon becomes larger. \ The limit $a_{\xi}\rightarrow0\,\ $corresponds to an
inertial frame. \ 

We imagine our box of thermal radiation as extending from a lower edge $\xi=a$
to an upper edge $\xi=b,$ and we introduce a spatial coordinate $\overline
{x}=\xi-a,$ the time $c\overline{t}=a\eta,$ and the length of the box $l=b-a.$
\ Then the normal mode in the Rindler frame given in Eq. (63) can be rewritten
for box length $l<<a$ as
\begin{align}
\varphi_{n}(\eta,\xi)  &  =\phi(c\overline{t},\overline{x})=\nonumber\\
&  =\left(  \frac{2}{\ln[(a+l)/a]}\right)  ^{1/2}\sin\left[  \frac{n\pi}%
{\ln[(a+l)/a]}\ln\left(  \frac{a+\overline{x}}{a}\right)  \right]  \cos\left[
\frac{n\pi}{\ln[(a+l)/a]}\frac{\overline{t}}{a}+\theta_{n}\right] \nonumber\\
&  \approx\left(  \frac{2a}{l}\right)  ^{1/2}\sin\left[  \frac{n\pi a}{l}%
\frac{\overline{x}}{a}\right]  \cos\left[  \frac{n\pi a}{l}\frac{\overline{t}%
}{a}+\theta_{n}\right] \nonumber\\
&  =a^{1/2}\left(  \frac{2}{l}\right)  ^{1/2}\sin\left[  \frac{n\pi
\overline{x}}{l}\right]  \cos\left[  \frac{n\pi\overline{t}}{l}+\theta
_{n}\right]
\end{align}
where we have used the approximation $\ln(1+l/a)\approx l/a$ which is valid
for large $a$, $a>>l.$ \ The over-all factor of $a^{1/2}$ is compensated by
the factor of $1/\xi=1/(a+\overline{x})\approx1/a$ which arises in the
normalization of the normal modes as seen in Eq. (65). \ Thus as the box of
radiation considered is further and further away from the event horizon, the
radiation normal modes go over to the normal modes (18) for a box in an
inertial frame. \ 

Next we consider the correlation function (73) in the limit of great distance
from the event horizon; this function becomes%
\begin{align}
&  <\varphi_{\alpha}(\eta,\xi)\frac{\partial}{\partial\eta^{\prime}}%
\varphi_{\alpha}(\eta^{\prime},\xi^{\prime})>_{\xi^{\prime}=\xi}=<\phi
_{\alpha}(c\overline{t},\overline{x})a\frac{\partial}{\partial(c\overline{t}%
)}\phi_{\alpha}(c\overline{t}^{\prime},\overline{x}^{\prime})>_{\overline
{x}=\overline{x}^{\prime}}\nonumber\\
&  =\alpha\hbar c\coth\left(  \alpha\frac{\eta-\eta^{\prime}}{2}\right)
=\alpha\hbar c\coth\left(  \alpha\frac{c(\overline{t}-\overline{t}^{\prime}%
)}{2a}\right)
\end{align}
Dividing Eq. (77) through by the factor $a$ (corresponding to the distance of
the lower end of the box from the event horizon), we have%
\begin{equation}
\left\langle \phi_{\alpha}(c\overline{t},\overline{x})\frac{\partial}%
{\partial(c\overline{t})}\phi_{\alpha}(c\overline{t}^{\prime},\overline
{x}^{\prime})\right\rangle _{\overline{x}=\overline{x}^{\prime}}=\frac{\alpha
}{a}\hbar c\coth\left(  \frac{\alpha}{a}\frac{c(\overline{t}-\overline
{t}^{\prime})}{2}\right)
\end{equation}
Now if, for fixed Rindler thermal parameter $\alpha,$ we take the limit as the
distance $a$ of the lower end of the box becomes increasingly far from the
event horizon, then we find from the power series expansion $\coth
z=1/z+z/3-...$ that
\begin{align}
&  <\phi_{\alpha}(c\overline{t},\overline{x})\frac{\partial}{\partial
(c\overline{t})}\phi_{\alpha}(c\overline{t}^{\prime},\overline{x}^{\prime
})>_{\overline{x}=\overline{x}^{\prime}}=\frac{\alpha}{a}\hbar c\left(
\frac{a}{\alpha}\frac{2}{c(\overline{t}-\overline{t}^{\prime})}+\frac{1}%
{3}\frac{\alpha}{a}c\frac{(\overline{t}-\overline{t}^{\prime})}{2}+...\right)
\nonumber\\
&  =2\hbar c\frac{1}{c(\overline{t}-\overline{t}^{\prime})}\text{ \ for the
limit \ }a\rightarrow\infty
\end{align}
Thus for fixed Rindler thermal parameter $\alpha,$ we simply recover Eq. (52)
(the zero-point radiation spectrum in an inertial frame) in the limit
$a\rightarrow\infty$ corresponding to being far from the event horizon.
\ However, from the Tolman-Ehrenfest relation (60), the temperature $T$ in the
box in the Rindler frame falls off with distance $a$ as $T=const/a$. \ Thus as
we consider a box whose lower edge $a$ is increasingly far from the event
horizon, the temperature indeed becomes ever smaller and goes to zero in the
limit of infinite distance. \ 

If we want to carry a box of thermal radiation at non-zero temperature $T$
back to an inertial frame, then we must keep on increasing the Rindler thermal
parameter $\alpha$ as the box considered is moved ever further from the event
horizon. \ If we compare the expression (77) for the correlation function in a
Rindler frame but very far from the event horizon with the correlation
function (51) for thermal radiation in an inertial frame, we see that the
ratio $\alpha/a$ must be proportional to the temperature $T$ at the lower end
of the box$,$ which temperature we wish to keep fixed. \ Accordingly, the
unknown function $TF[Tc(t-t^{\prime})]$ in the inertial frame arising from the
correlation function in Eq. (51) is given by
\begin{equation}
<\phi_{T}(c\overline{t},\overline{x})\partial_{c\overline{t}}\phi
_{T}(c\overline{t}^{\prime},\overline{x}^{\prime})>_{\overline{x}=\overline
{x}^{\prime}}=TF[Tc(t-t^{\prime})]=\frac{\alpha}{a}\hbar c\coth\left(
\frac{\alpha}{a}\frac{c(\overline{t}-\overline{t}^{\prime})}{2}\right)
\end{equation}
where there is some constant of proportionality between the temperature $T$
and the ratio $\alpha/a$.

The spectral function for thermal radiation in an inertial frame
$f^{2}(|k|,T)|k|=g(|k|/T)$ can be found from Eq. (80) by the same inverse
Fourier sine transform as was evaluated in Eq. (75), and so we finally obtain
the spectrum for classical scalar thermal radiation in an inertial frame%
\begin{align}
f^{2}(|k|,T)|k|  &  =g(|k|/T)=4\int_{0}^{\infty}dtc\left[  \frac{\alpha}%
{a}\hbar c\coth\left(  \frac{\alpha}{a}\frac{ct}{2}\right)  \right]
\sin(|k|ct)\nonumber\\
&  =4\pi\hbar c\coth\left(  \frac{|k|\pi}{\alpha/a}\right)
\end{align}
The proportionality constant between $T$ and $\alpha/a$ can be fixed by the
limits of Eqs. (46) and (47) requiring that the energy $U(k)$ of a normal mode
become $k_{B}T$ \ for fixed wave number $k$ and high temperature $T.$ \ Then
we find
\begin{equation}
f^{2}(|k|,T)|k|=g(|k|/T)=4\pi\hbar c\coth\left(  \frac{\hbar c|k|}{2k_{B}%
T}\right)
\end{equation}
and an energy spectrum corresponding to an energy per normal mode at wave
number $k$ with $\omega=c|k|,$%
\begin{align}
U(k)  &  =\frac{1}{8\pi}f^{2}(|k|,T)k^{2}=\frac{1}{2}\hbar c|k|\coth\left(
\frac{\hbar c|k|}{2k_{B}T}\right) \nonumber\\
&  =\frac{1}{2}\hbar\omega\coth\left(  \frac{\hbar\omega}{2k_{B}T}\right)
=\frac{1}{2}\hbar\omega+\frac{1}{\exp(\hbar\omega/k_{B}T)-1}%
\end{align}
This result is just the Planck spectrum of thermal radiation including
zero-point radiation.\ \ Thus we find that by considering the box of thermal
radiation ever farther from the event horizon while keeping the temperature
inside the box fixed, the coordinates become ever closer to inertial
coordinates, the normal modes go over to inertial-frame normal modes, and the
spectrum becomes the thermal radiation spectrum in an inertial frame. \ Thus
we have successfully obtained the Planck spectrum of thermal radiation from
the use of zero-point radiation and the information of a relativistic Rindler frame.

What is thermal radiation? \ Apparently thermal radiation is time-stationary
radiation in an inertial frame which follows the same basic spectral form as
Lorentz-invariant zero-point radiation takes in a Rindler frame.

\section{Discussion \ }

All the textbooks\cite{Eisberg} of physics claim that classical physics is
incapable of obtaining the Planck spectrum of thermal radiation and that one
must use the quantum statical mechanics of photons in order to account for
thermal radiation. However, the usual classical analysis applies
nonrelativistic classical statistical mechanics\cite{ER12} to the radiation
modes of the relativistic radiation field or else treats the scattering of the
relativistic radiation field by nonrelativistic mechanical
scatterers.\cite{scatt} \ These calculations involve inconsistent mixtures of
Galilean-invariant and Lorentz-invariant physics. \ The present analysis
obtains the Planck spectrum within a consistent relativistic analysis. \ Again
it is suggested that only a fully relativistic treatment of relativistic
radiation is consistent with classical radiation equilibrium. \ This point of
view agrees with the scattering result reported for electromagnetic radiation
that classical zero-point radiation remains zero-point radiation under
scattering by a relativistic charged particle in a Coulomb
potential.\cite{FOP}

\section{Acknowledgement}

The present analysis is related to work in quantum field theory\cite{Davies}%
\cite{Crispino} which deals with the "thermal effects of
acceleration."\cite{connections} \ It was first within the context of the
quantum field theory that there arose a connection between thermal radiation
and acceleration. \ Work from the quantum calculations has provided hints for
the classical analysis presented here. \

\bigskip

\end{document}